\documentstyle{article}

\title{Quantum Mechanics as a Classical Theory IV:\\
The Negative Mass Conjecture}
\author{L. S. F. Olavo\\
Departamento de Fisica - Universidade de Brasilia - UnB\\
70910-900 - Brasilia - D.F.- Brazil}

\begin{document}

\maketitle
\begin{abstract}
The following two papers form a natural development of a previous series of
three articles on the foundations of quantum mechanics; they are intended to
take the theory there developed to its utmost logical and epistemological
consequences. We show in the first paper that relativistic quantum mechanics
might accommodate without ambiguities the notion of negative masses. To
achieve this, we rewrite all of its formalism for integer and half integer
spin particles and present the world revealed by this conjecture. We also
base the theory on the second order Klein-Gordon's and Dirac's equations and
show that they can be stated with only positive definite energies. In the
second paper we show that the general relativistic quantum mechanics derived
in paper II of this series supports this conjecture.
\end{abstract}

\section{General Introduction}

What is the job of a theoretical physicist? The first answer that comes to
us in a somewhat precipitate manner is: The theoretical physicist's job is
to say how the world is. Despite the obvious philosophical fragility of such
an assertion, hardly adjustable with the method of systematic doubt of
science, this answer gives us a key for a more adequate approach. The
theoretical physicist has not the mission of saying how the world is but,
rather, the job to explain how the world might be. Only the experiments have
the final word about, among all the numerous possible worlds furnished by a
theory, which one is more adequate. The history of science of the last four
centuries has shown that we shall not underestimate any of the models we
uncover with our interpretations of the underlying formal apparatus.

This is the spirit underlying the first paper of this series. In this paper,
we will show that relativistic quantum mechanics admits an interpretation
very different from the one usually accepted. Since we are not interested in
obtaining a new relativistic quantum mechanical formalism, its formal
apparatus will be kept intact. Our interest is to uncover another world,
equally allowed by this apparatus, and show that an arbitrary choice has
hidden this world.

The methodological criterion that one should apply in judging the merits of
this work cannot be related to its applicability, since we keep the pure
formal apparatus intact and expect the same formal outcomes. This criterion
has to do with the world picture that emerges from one and the other
theoretical interpretations. It is from this point that they become
different theories and expect answers from Nature so distinct as excluding.
The corroboration of one choice or the other is left, however, for the
experiments...

In the second section of this paper, we will introduce, in a rather
intuitive way, the main idea of this first paper. We will claim that the
relativistic quantum mechanical formalism can accommodate a world with
negative masses. We will use the Klein-Gordon theory in elaborating such an
argument.

In the third section, we will develop the considerations made in the
previous one into a more mathematical format.

We will make, in the fourth section, an extension of the previous formalism
to apply it to particles with half integral spin. We will then use the
second order Dirac's equation.

After the fourth section, the first part of this series of two papers will
be complete. We will have demonstrated that all relativistic quantum
mechanics can be rewritten to accommodate negative masses. We then make our
conclusions.

We devote the appendix to show that relativistic quantum mechanics based on
second order equations can be rewritten to admit only positive probability
densities. We then show that the solution of this problem bears some
resemblance with the formalism developed in the main text.

The next paper is a continuation of the first. We make an application of the
general relativistic quantum mechanical formalism already derived\cite{1,2,3}
to a simple, but highly instructive, example.

In the second section of that paper, we apply the formalism to the simple
problem of a test mass gravitating around a heavy body (we call this problem
the quantum Schwartzchild problem). From the results so obtained, we show
that this general relativistic quantum theory supports the negative mass
conjecture of the first paper.

In the third section we make our final conclusions.

\section{Introduction}

When the Klein-Gordon theory (hereafter KG) was proposed, the possibility of
negative probability densities was one of its main deficiencies. The
solution met was to multiply this density by the modulus of the electric
charge and to consider it as a charge, rather than a probability, density.
This attitude, however, seems to be based on an arbitrary choice that has
hidden other possibilities.

The usual interpretation of the relativistic quantum mechanical formalism
assumes, by principle, that there can be no negative masses in Nature\cite{%
4,5}. We now turn to show that we can eliminate this constraint from
the interpretation without incurring into inconsistencies.

In a previous series of papers, we have shown that the KG probability
density, defined as
\begin{equation}
\label{1}j_0\left( x\right) =\frac{i\hbar }{2mc}\left( \phi ^{*}\left(
x\right) \partial _0\phi \left( x\right) -\phi \left( x\right) \partial
_0\phi ^{*}\left( x\right) \right) ,
\end{equation}
where $m$ is the particle's mass and $\phi $ is the associated probability
amplitude, does not require to be multiplied by any charge to represent a
true probability density if we accept that we should have negative masses
for antiparticles. We then could write the probability density as
\begin{equation}
\label{2}\rho _\lambda \left( x\right) =\frac{i\hbar }{2\lambda mc}\left[
\phi ^{*}\left( x\right) \partial _0\phi \left( x\right) -\phi \left(
x\right) \partial _0\phi ^{*}\left( x\right) \right] _{+\ },
\end{equation}
where $[\ ]_{+}$ implies that we take only the positive signal of the
quantity inside brackets, and the parameter $\lambda $ defines if the
density refers to particles or antiparticles:
\begin{equation}
\label{3}\lambda =sign\left( \phi ^{*}\left( x\right) \partial _0\phi \left(
x\right) -\phi \left( x\right) \partial _0\phi ^{*}\left( x\right) \right)
=\left\{
\begin{array}{c}
+1\ \ \ \ \ \quad
\mbox{for\ particles} \\ -1\quad \mbox{for\ antiparticles}
\end{array}
\right. .
\end{equation}
We might thus interpret a negative probability density as a positive one
describing negative mass particles (antiparticles). In such a case, the mass
distribution can be written as:
\begin{equation}
\label{4}\rho _\lambda ^{mass}\left( x\right) =m\rho _\lambda \left(
x\right) =\frac{i\hbar }{2\lambda c}\left[ \phi ^{*}\left( x\right) \partial
_0\phi \left( x\right) -\phi \left( x\right) \partial _0\phi ^{*}\left(
x\right) \right] _{+}
\end{equation}
and will be positive for particles and negative for antiparticles. From the
very definition of the parameter $\lambda $ it is easy to see that the
complex conjugation, defined by $\phi \rightarrow \phi ^{*}$, implies $%
\lambda \rightarrow -\lambda $ and thus, in the mapping of particles into
antiparticles.

We can handle with electromagnetic fields in a way similar to the usually
done in the literature. We then have%
$$
\rho _\lambda \left( x\right) =\frac{i\hbar }{2\lambda m_0c}\left[ \phi
^{*}\left( x\right) \partial _0\phi \left( x\right) -\phi \left( x\right)
\partial _0\phi ^{*}\left( x\right) \right] _{+}-\frac{2e}{m_0c^2}\Phi
\left( x\right) \phi ^{*}\left( x\right) \phi \left( x\right) =
$$
\begin{equation}
\label{5}=\frac{i\hbar }{2\lambda mc}\left[ \phi ^{*}\left( x\right)
\partial _0\phi \left( x\right) -\phi \left( x\right) \partial _0\phi
^{*}\left( x\right) \right] _{+}-\frac{2\lambda e}{\lambda mc^2}\Phi \left(
x\right) \phi ^{*}\left( x\right) \phi \left( x\right) ,
\end{equation}
where $\Phi $ is the scalar electromagnetic potential. Collecting terms we
get
$$
\label{6}\rho _\lambda \left( x\right) =
$$
\begin{equation}
=\frac 1{2\lambda mc^2}\left\{ \left[
\phi ^{*}\left( x\right) i\hbar \frac{\partial \phi \left( x\right) }{%
\partial t}-\phi \left( x\right) i\hbar \frac{\partial \phi ^{*}\left(
x\right) }{\partial t}\right] _{+}-2\lambda e\Phi \left( x\right) \phi
^{*}\left( x\right) \phi \left( x\right) \right\} ,
\end{equation}
which shows that the complex conjugation of the amplitudes also implies in
the change of the electric charge (we can also see this looking directly at
the KG equation). We conclude that, in the present theory, the complex
conjugation operation has the effect of changing the mass and charge signs.
This implies that particles shall have these properties with the opposite
sign of the associated antiparticles.

We know from the experiments that particle-antiparticle pairs, when
subjected to homogeneous magnetic fields, move along opposite circular
trajectories; this is the reason for the usual interpretation considering
the charge of particles and antiparticles to have opposite sign. In the
present interpretation, both mass and charge change sign; thus, the ratio $%
e/m$ does not change its sign.

It is important to stress, however, that the charge and the mass appear in
the expression for the trajectory of the pair together with the velocity of
its components. We now turn to see what happens with these velocities in our
formalism.

Let us consider then the free particle-antiparticle solutions:
\begin{equation}
\label{7}\phi _\lambda \left( x\right) =\exp \left[ -\lambda i\left( E_pt-%
{\bf p\cdot r}\right) /\hbar \right] ,
\end{equation}
where the evolution parameter $E_{p\mbox{ }}$and the momentum ${\bf p}$ are
given by
\begin{equation}
\label{8}E_p=mc^2/\sqrt{1-v^2/c^2}\quad ;\quad {\bf p}=m{\bf v}/\sqrt{%
1-v^2/c^2}.
\end{equation}
The probability density and flux, in the absence of electromagnetic fields,
are given by
\begin{equation}
\label{9}\rho _\lambda \left( x\right) =\frac{E_p}{\lambda mc^2}\quad ;\quad
{\bf j}_\lambda \left( x\right) =\frac{{\bf p}}{\lambda mc}.
\end{equation}
If we put
\begin{equation}
\label{10}\frac{{\bf p}}m={\bf v}\quad \Rightarrow \left\{
\begin{array}{c}
{\bf p}_a=\left( -m\right) {\bf v}_a \\ {\bf p}_p=\left( +m\right) {\bf v}_p
\end{array}
\right. \quad ,
\end{equation}
where ${\bf v}_p$ and ${\bf v}_a$ are the velocities of the particle and
antiparticle, respectively, and ${\bf p}_p$, ${\bf p}_a$ their momenta, we
then get
\begin{equation}
\label{11}{\bf j}_\lambda \left( x\right) =\lambda \frac{{\bf v}}c,
\end{equation}
which can be interpreted as meaning that the flux of particles in one
direction is equivalent to the flux of antiparticles in the opposite
direction. In this manner, we expect that, when gravitational forces are
present, particles and antiparticles behave in the way shown in figure 1.
These forces obviously do not pertain to the framework of the present
theory; a theory that takes gravitation into account will be dealt with in
the next paper. However it is noteworthy that particles and antiparticles
will not respond perversely to homogeneous magnetic fields as one could in
principle think\cite{6}. Indeed, when electromagnetic fields are present
and taking $+e$ and $-e$ as the particle's and the antiparticle's charges,
respectively, we have:
\begin{equation}
\label{12}\rho _\lambda \left( x\right) =\frac 1{\lambda mc^2}\left(
E_p-\lambda e\Phi \right) \quad ;\quad {\bf j}_\lambda \left( x\right)
=\frac 1{\lambda mc}\left( {\bf p-}\lambda \frac ec{\bf A}\right) ,
\end{equation}
which gives, for the flux
\begin{equation}
\label{13}{\bf j}_\lambda \left( x\right) =\frac 1c\left( \lambda {\bf v-}%
\frac e{mc}{\bf A}\right) .
\end{equation}

This shows that particles and antiparticles have velocity vectors with
opposite signs compared to the potential vector. This property is sufficient
to explain their behavior under the influence of a homogeneous magnetic
field (figure 2).

We now proceed, in the next two sections, to rewrite the mathematical
apparatus to state formally our conjecture.

\section{Klein-Gordon's Theory with Negative Mass}

If we depart from the hypothesis that Nature can reveal entities with masses
of both signs, we then expect to find in It all the combinations shown in
table I. We might use the Feshbach-Villars decomposition to relate all the
possibilities furnished by nature with the KG equation. By means of this
decomposition, the KG equation
\begin{equation}
\label{14}\frac 1{c^2}\left( i\hbar \frac \partial {\partial t}-e\Phi
\right) ^2\varphi =\frac 1{2m}\left( \frac \hbar i\nabla -\frac ec{\bf A}%
\right) ^2\varphi +m^2c^2\varphi ,
\end{equation}
when we use
\begin{equation}
\label{15}\varphi _0\left( {\bf r},t\right) =\left[ \frac \partial {\partial
t}+\frac{ie}\hbar \Phi \left( {\bf r},t\right) \right] \varphi \left( {\bf r}%
,t\right) ,
\end{equation}
and
\begin{equation}
\label{16}\varphi _1=\frac 12\left[ \varphi _0+\frac{i\hbar }{m_0c^2}\varphi
\right] \quad ;\quad \varphi _2=\frac 12\left[ \varphi _0-\frac{i\hbar }{%
m_0c^2}\varphi \right] ,
\end{equation}
becomes the following system of equations
\begin{equation}
\label{17}\left[ i\hbar \frac \partial {\partial t}-e\Phi \right] \varphi
_1=\frac 1{2m}\left[ \frac \hbar i\nabla -\frac ec{\bf A}\right] ^2\left(
\varphi _1+\varphi _2\right) +mc^2\varphi _1;
\end{equation}
\begin{equation}
\label{18}\left[ i\hbar \frac \partial {\partial t}-e\Phi \right] \varphi
_2= \frac{-1}{2m}\left[ \frac \hbar i\nabla -\frac ec{\bf A}\right] ^2\left(
\varphi _1+\varphi _2\right) -mc^2\varphi _2;
\end{equation}
together with their complex conjugate
\begin{equation}
\label{19}\left[ i\hbar \frac \partial {\partial t}+e\Phi \right] \varphi
_1^{*}=\frac{-1}{2m}\left[ \frac \hbar i\nabla +\frac ec{\bf A}\right]
^2\left( \varphi _1^{*}+\varphi _2^{*}\right) -mc^2\varphi _1^{*};
\end{equation}
\begin{equation}
\label{20}\left[ i\hbar \frac \partial {\partial t}+e\Phi \right] \varphi
_2^{*}=\frac 1{2m}\left[ \frac \hbar i\nabla +\frac ec{\bf A}\right]
^2\left( \varphi _1^{*}+\varphi _2^{*}\right) +mc^2\varphi _2^{*}.
\end{equation}

With the notation above we note that it is possible to make a connection
between the amplitudes and the particles signs of mass and charge they
represent
\begin{equation}
\label{21}\varphi _1\Longleftrightarrow \left( +,+\right) \ ;\ \varphi
_2\Longleftrightarrow \left( -,+\right) ,
\end{equation}
\begin{equation}
\label{22}\varphi _1^{*}\Longleftrightarrow \left( -,-\right) \ ;\ \varphi
_2^{*}\Longleftrightarrow \left( +,-\right) ,
\end{equation}
where $(A,B)$ represents an entity with mass and charge signs $A$ and $B$,
respectively.

We made the {\it choice}, in the last section, to represent antiparticles
with the signs of the mass and charge reverted as related to the particle.
In agreement with this choice, we shall attribute for pairs of such entities
an amplitude and its complex conjugate, as become clear from equations (\ref
{17}-\ref{20}).

We can now define the two-component spinors
\begin{equation}
\label{23}\Psi =\left(
\begin{array}{c}
\varphi _1 \\
\varphi _2
\end{array}
\right) \quad ;\quad \Psi ^{*}=\left(
\begin{array}{c}
\varphi _1^{*} \\
\varphi _2^{*}
\end{array}
\right) ,
\end{equation}
together with the Pauli matrices
\begin{equation}
\label{24}\sigma _1=\left(
\begin{array}{cc}
0 & 1 \\
1 & 0
\end{array}
\right) \ ;\ \sigma _2=\left(
\begin{array}{cc}
0 & -i \\
i & 0
\end{array}
\right) \ ;\ \sigma _3=\left(
\begin{array}{cc}
1 & 0 \\
0 & -1
\end{array}
\right)
\end{equation}
and rewrite the system (17-20) as
\begin{equation}
\label{25}\left( i\hbar \frac \partial {\partial t}-e\Phi \right) \Psi
=\left[ \frac 1{2m}\left( \frac \hbar i\nabla -\frac ec{\bf A}\right)
^2\left( \sigma _3+i\sigma _2\right) +mc^2\sigma _3\right] \Psi ,
\end{equation}
or else
\begin{equation}
\label{26}\Psi _{c1}\left( i\hbar \frac \partial {\partial t}+e\Phi \right)
=\Psi _{c1}\left[ \frac{-1}{2m}\left( \frac \hbar i\nabla +\frac ec{\bf A}%
\right) ^2\left( \sigma _3+i\sigma _2\right) -mc^2\sigma _3\right] ,
\end{equation}
where
\begin{equation}
\label{27}\Psi _{c1}=\Psi ^{\dagger }\sigma _3.
\end{equation}

Using the basis
\begin{equation}
\label{28}{\bf e}_1=\left(
\begin{array}{c}
1 \\
0
\end{array}
\right) \ ;\ {\bf e}_2=\left(
\begin{array}{c}
0 \\
1
\end{array}
\right) ,
\end{equation}
we can adopt the convention
\begin{equation}
\label{29}u_0^{\left( P,+\right) },
\end{equation}
where the index zero indicates that we are in the inertial frame of
reference and the dyad $(P,+)$ implies that the related spinor describe a
particle with positive charge. It is then possible to write the four
possible functions (Table 1) as
\begin{equation}
\label{30}u_0^{\left( P,+\right) }={\bf e}_1e^{-iE_p\tau /\hbar }\ ;\
u_0^{\left( A,-\right) }={\bf e}_1e^{+iE_p\tau /\hbar };
\end{equation}
\begin{equation}
\label{31}u_0^{\left( A,+\right) }={\bf e}_2e^{+iE_p\tau /\hbar }\ ;\
u_0^{\left( P,-\right) }={\bf e}_2e^{-iE_p\tau /\hbar };
\end{equation}
where
\begin{equation}
\label{32}E_p=mc^2.
\end{equation}

We might still define a charge conjugation by the operation
\begin{equation}
\label{33}\Psi _c=\sigma _1\Psi ^{*},
\end{equation}
that satisfies a KG equation with the same mass sign but with the charge
sign reverted. This spinor, however, cannot be now a candidate to represent
antiparticles related to $\Psi $ since only the charge sign is reverted.

We note, however, that the difference between complex conjugation and charge
conjugation is relevant only in the realm of a theory that distinguishes
mass signs. We can see this by covering the mass column in table I or II and
noting that, in this case, those amplitudes are degenerate.

The probability density can be immediately obtained and is given by
\begin{equation}
\label{34}\rho =\Psi ^{\dagger }\sigma _3\Psi =\Psi _{c1}\Psi ,
\end{equation}
where now, when permuting the amplitudes, we keep the sign. This should
happen because each amplitude has a component related to a particle and
another to an antiparticle (with the same sign of the charge).

The current or flux density can be easily obtained and is given by
\begin{equation}
\label{35}{\bf j}=\frac 1{2m}\left[ \Psi _{c1}\Lambda \nabla \Psi -\left(
\nabla \Psi _{c1}\right) \Lambda \Psi \right] -\frac{e\hbar }{mc}{\bf A}\Psi
_{c1}\Lambda \Psi ,
\end{equation}
where
\begin{equation}
\label{36}\Lambda =\left( \sigma _3+i\sigma _2\right) .
\end{equation}

Before we go on with the study of particles with spin, it is interesting to
consider particles with null charge. The usual interpretation denies these
particles of being described by the KG formalism (at least if there is no
interaction capable of distinguishing them). This is the case, for example,
of the pion zero. Being a null charge particle, the associated charge
density must be identically zero. These particles are then said to be their
own antiparticles. We cannot say this in the present theory. Here, the pion
zero might manifest itself with two masses of different signs that can be
distinguished by a gravitational field. We are then faced with a pion zero
and an antipion zero.

In the next section, we continue developing an analogous theory for half-
spin particles.

\section{Dirac's Theory with Negative Mass}

We wish to develop a similar formalism for Dirac's equation as was done for
Klein-Gordon's. As was already mentioned in the first papers of this
series\cite{1,2,3}, we shall consider the second order Dirac's equation as
the fundamental one rather than the first order equation.

We then depart from Dirac's second order equation
$$
\frac 1{c^2}\left( i\hbar \frac \partial {\partial t}-e\Phi \right) ^2\left(
\begin{array}{c}
\varphi \\
\chi
\end{array}
\right) =\left[ \left( \frac \hbar i\nabla -\frac ec{\bf A}\right) ^2{\bf 1}%
+m^2c^2{\bf 1}+\right.
$$
\begin{equation}
\label{37}\left. +\frac{e\hbar }c\left(
\begin{array}{cc}
\sigma \cdot {\bf H} & {\bf 0} \\ {\bf 0} & \sigma \cdot {\bf H}
\end{array}
\right) -i\frac{e\hbar }c\left(
\begin{array}{cc}
{\bf 0} & \sigma \cdot
{\bf E} \\ \sigma \cdot {\bf E} & {\bf 0}
\end{array}
\right) \right] \left(
\begin{array}{c}
\varphi \\
\chi
\end{array}
\right)
\end{equation}
where $\varphi $, $\chi $ are two-component spinors, while ${\bf H}$ and $%
{\bf E}$ are the magnetic and electric field, respectively.

The expression for the probability density can be easily obtained and is
given by
\begin{equation}
\label{38}\rho _\lambda =\frac 1{2\lambda mc^2}\left\{ \left[ \psi ^{\dagger
}\beta ih\frac{\partial \psi }{\partial t}-\left( i\hbar \frac{\partial \psi
^{\dagger }}{\partial t}\right) \beta \psi \right] _{+}-2\lambda e\Phi \psi
^{\dagger }\beta \psi \right\} ,
\end{equation}
where $\psi $ is the four-component spinor
\begin{equation}
\label{38a}\psi =\left(
\begin{array}{c}
\varphi \\
\chi
\end{array}
\right)
\end{equation}
and $\beta $ is the usual spin parity operator in Dirac's representation. We
are now in position to rewrite Dirac's formalism in the format given in the
previous section. We will thus use a simile of the Feshbach-Villars
decomposition applied to Dirac's second order equation.

Such a decomposition is attained if we define
\begin{equation}
\label{39}\varphi _0=\left[ \frac \partial {\partial t}+\frac{ie}\hbar \Phi
\right] \varphi \ ;\ \chi _0=\left[ \frac \partial {\partial t}+\frac{ie}%
\hbar \Phi \right] \chi
\end{equation}
and
\begin{equation}
\label{40}\left\{
\begin{array}{c}
\varphi _1=\frac 12\left( \varphi _0+
\frac{i\hbar }{mc}\varphi \right) \\ \varphi _2=\frac 12\left( \varphi _0-
\frac{i\hbar }{mc}\varphi \right)
\end{array}
\right. \ ;\ \left\{
\begin{array}{c}
\chi _1=\frac 12\left( \chi _0+
\frac{i\hbar }{mc}\chi \right) \\ \chi _2=\frac 12\left( \chi _0-\frac{%
i\hbar }{mc}\chi \right)
\end{array}
\right. ,
\end{equation}
where $\varphi _1$, $\varphi _2$, $\chi _1$, $\chi _2$ are two-component
spinors. We are then led to the following equations:
$$
\left( i\hbar \frac \partial {\partial t}-e\Phi \right) \varphi _1=\left[
\frac 1{2m}\left( \frac \hbar i\nabla -\frac ec{\bf A}\right) ^2{\bf 1}-
\frac{e\hbar }{2mc}\sigma \cdot {\bf H}\right] \left( \varphi _1+\varphi
_2\right) +
$$
\begin{equation}
\label{41}+mc^2\varphi _1+\frac{ie\hbar }{2mc}\sigma \cdot {\bf E}\left(
\chi _1+\chi _2\right) ;
\end{equation}
$$
\left( i\hbar \frac \partial {\partial t}-e\Phi \right) \varphi _2=\left[
\frac{-1}{2m}\left( \frac \hbar i\nabla -\frac ec{\bf A}\right) ^2{\bf 1+}
\frac{e\hbar }{2mc}\sigma \cdot {\bf H}\right] \left( \varphi _1+\varphi
_2\right) -
$$
\begin{equation}
\label{42}-mc^2\varphi _1-\frac{ie\hbar }{2mc}\sigma \cdot {\bf E}\left(
\chi _1+\chi _2\right) ;
\end{equation}
$$
\left( i\hbar \frac \partial {\partial t}-e\Phi \right) \chi _1=\left[ \frac
1{2m}\left( \frac \hbar i\nabla -\frac ec{\bf A}\right) ^2{\bf 1}-\frac{%
e\hbar }{2mc}\sigma \cdot {\bf H}\right] \left( \chi _1+\chi _2\right) +
$$
\begin{equation}
\label{43}+mc^2\chi _1+\frac{ie\hbar }{2mc}\sigma \cdot {\bf E}\left(
\varphi _1+\varphi _2\right) ;
\end{equation}
$$
\left( i\hbar \frac \partial {\partial t}-e\Phi \right) \chi _2=\left[ \frac{%
-1}{2m}\left( \frac \hbar i\nabla -\frac ec{\bf A}\right) ^2{\bf 1+}\frac{%
e\hbar }{2mc}\sigma \cdot {\bf H}\right] \left( \chi _1+\chi _2\right) -
$$
\begin{equation}
\label{44}-mc^2\chi _1-\frac{ie\hbar }{2mc}\sigma \cdot {\bf E}\left(
\varphi _1+\varphi _2\right) .
\end{equation}
These equations, togheter with their complex conjugate, cover all the
possibilities we expect from Nature when leting for the existence of
negative masses (Table 2).

Defining the eight-component spinor
\begin{equation}
\label{45}\Psi =\left[
\begin{array}{c}
\varphi _1 \\
\varphi _2 \\
\chi _1 \\
\chi _2
\end{array}
\right] \ ;\ \varphi _i=\left[
\begin{array}{c}
\varphi _{i1} \\
\varphi _{i2}
\end{array}
\right] \ ;\ \chi _i=\left[
\begin{array}{c}
\chi _{i1} \\
\chi _{i2}
\end{array}
\right] \ ,
\end{equation}
the matrices
\begin{equation}
\label{46}\Sigma _1=\left[
\begin{array}{cccc}
{\bf 0} & {\bf +1} & {\bf 0} & {\bf 0} \\ {\bf +1} & {\bf 0} & {\bf 0} &
{\bf 0} \\ {\bf 0} & {\bf 0} & {\bf 0} & {\bf +1} \\ {\bf 0} & {\bf 0} &
{\bf +1} & {\bf 0}
\end{array}
\right] \ ;\ \Sigma _2=\left[
\begin{array}{cccc}
{\bf 0} & -{\bf i} & {\bf 0} & {\bf 0} \\ {\bf i} & {\bf 0} & {\bf 0} & {\bf %
0} \\ {\bf 0} & {\bf 0} & {\bf 0} & -
{\bf i} \\ {\bf 0} & {\bf 0} & {\bf i} & {\bf 0}
\end{array}
\right]
\end{equation}
\begin{equation}
\label{47}\Sigma _3=\left[
\begin{array}{cccc}
+{\bf 1} & {\bf 0} & {\bf 0} & {\bf 0} \\ {\bf 0} & {\bf -1} & {\bf 0} &
{\bf 0} \\ {\bf 0} & {\bf 0} & +{\bf 1} & {\bf 0} \\ {\bf 0} & {\bf 0} &
{\bf 0} & {\bf -1}
\end{array}
\right] \ ;\ \alpha _1=\left[
\begin{array}{cccc}
{\bf 0} & {\bf 0} & {\bf 0} & {\bf 1} \\ {\bf 0} & {\bf 0} & {\bf 1} & {\bf 0%
} \\ {\bf 0} & {\bf 1} & {\bf 0} & {\bf 0} \\ {\bf 1} & {\bf 0} & {\bf 0} &
{\bf 0}
\end{array}
\right] ;
\end{equation}
\begin{equation}
\label{48}\alpha _2=\left[
\begin{array}{cccc}
{\bf 0} & {\bf 0} & {\bf 0} & {\bf -i} \\ {\bf 0} & {\bf 0} & {\bf +i} &
{\bf 0} \\ {\bf 0} & {\bf -i} & {\bf 0} & {\bf 0} \\ {\bf +i} & {\bf 0} &
{\bf 0} & {\bf 0}
\end{array}
\right] \ ;\ \alpha _3=\left[
\begin{array}{cccc}
{\bf 0} & {\bf 0} & {\bf +1} & {\bf 0} \\ {\bf 0} & {\bf 0} & {\bf 0} & {\bf %
-1} \\ {\bf +1} & {\bf 0} & {\bf 0} & {\bf 0} \\ {\bf 0} & {\bf -1} & {\bf 0}
& {\bf 0}
\end{array}
\right]
\end{equation}
and
\begin{equation}
\label{49}\beta =\left[
\begin{array}{cccc}
{\bf +1} & {\bf 0} & {\bf 0} & {\bf 0} \\ {\bf 0} & {\bf +1} & {\bf 0} &
{\bf 0} \\ {\bf 0} & {\bf 0} & {\bf -1} & {\bf 0} \\ {\bf 0} & {\bf 0} &
{\bf 0} & {\bf -1}
\end{array}
\right] ,
\end{equation}
where each element is a $2\times 2$ matrix, we can write the above system of
equations as:%
$$
\left( i\hbar \frac \partial {\partial t}-e\Phi \right) \Psi =\left[ \frac
1{2m}\left( \frac \hbar i\nabla -\frac ec{\bf A}\right) ^2{\bf 1}-\frac{%
e\hbar }{2mc}\sigma \cdot {\bf H}\right] \left( \Sigma _3+i\Sigma _2\right)
\Psi +
$$
\begin{equation}
\label{50}+mc^2\Sigma _3\Psi +\frac{ie\hbar }{2mc}\sigma \cdot {\bf E}\left(
\alpha _3+i\alpha _2\right) \Psi .
\end{equation}

It is then easy to show that
\begin{equation}
\label{51}\Psi _{c1}=i\beta \sigma _2\Psi ^{*}
\end{equation}
is a solution of%
$$
\left( i\hbar \frac \partial {\partial t}+e\Phi \right) \Psi _{c1}=\left[
\frac{-1}{2m}\left( \frac \hbar i\nabla +\frac ec{\bf A}\right) ^2{\bf 1}-
\frac{e\hbar }{2mc}\sigma \cdot {\bf H}\right] \left( \Sigma _3+i\Sigma
_2\right) \Psi _{c1}-
$$
\begin{equation}
\label{52}-mc^2\Sigma _3\Psi _{c1}+\frac{ie\hbar }{2mc}\sigma \cdot {\bf E}%
\left( \alpha _3+i\alpha _2\right) \Psi _{c1}
\end{equation}
which is the same equation solved by $\Psi $ with the signs of the mass and
the charge inverted, but with the same parity.

We can also show that
\begin{equation}
\label{53}\Psi _{c2}=\Psi ^{\dagger }\Sigma _3i\alpha _3\beta
\end{equation}
is a solution of%
$$
\Psi _{c2}\left( i\hbar \frac \partial {\partial t}+e\Phi \right) =\Psi
_{c2}\left( \Sigma _3+i\Sigma _2\right) \left[ \frac{-1}{2m}\left( \frac
\hbar i\nabla +\frac ec{\bf A}\right) ^2{\bf 1}-\frac{e\hbar }{2mc}\sigma
\cdot {\bf H}\right] -
$$
\begin{equation}
\label{54}-mc^2\Psi _{c2}\Sigma _3-\frac{ie\hbar }{2mc}\sigma \cdot {\bf E}%
\Psi _{c2}\left( \alpha _3+i\alpha _2\right)
\end{equation}
which is similar to the one solved by $\Psi $ with the signs of the mass,
the charge and the parity inverted, while keeping the signs of the spins.

Both the above amplitudes are candidates to represent antiparticles of $\Psi
$ since we have used, until now, only the criterion of the mass and charge
signs. We might write them explicitly as
\begin{equation}
\label{55}\Psi =\left[
\begin{array}{c}
\varphi _{11}(+,+) \\
\varphi _{12}(+,+) \\
\varphi _{21}(-,+) \\
\varphi _{22}(-,+) \\
\chi _{11}(+,+) \\
\chi _{12}(+,+) \\
\chi _{21}(-,+) \\
\chi _{22}(-,+)
\end{array}
\right] \ \Rightarrow \ \Psi _{c1}=\left[
\begin{array}{c}
+\varphi _{12}^{*}(-,-) \\
-\varphi _{11}^{*}(-,-) \\
+\varphi _{22}^{*}(+,-) \\
-\varphi _{21}^{*}(+,-) \\
-\chi _{12}^{*}(-,-) \\
+\chi _{11}^{*}(-,-) \\
-\chi _{22}^{*}(+,-) \\
+\chi _{21}^{*}(+,-)
\end{array}
\right] \ ;\ \Psi _{c2}=\left[
\begin{array}{c}
+\chi _{11}^{*}(-,-) \\
+\chi _{12}^{*}(-,-) \\
+\chi _{21}^{*}(+,-) \\
+\chi _{22}^{*}(+,-) \\
-\varphi _{11}^{*}(-,-) \\
-\varphi _{12}^{*}(-,-) \\
-\varphi _{21}^{*}(+,-) \\
-\varphi _{22}^{*}(+,-)
\end{array}
\right] ,
\end{equation}
where we also show, inside parenthesis, the signs of the mass and the charge
related to each component of the spinors (we took the transpose of the
line-spinor). This arrangement shows more clearly what relation particles
exhibit with antiparticles by means of the above mentioned functions.

We now define the element
\begin{equation}
\label{56}u_{0\uparrow (+)}^{(P,+)},
\end{equation}
as the eight-component spinor where: the index zero denotes that we are in
the rest frame, the up arrow indicates the spin up (upon action of operator $%
\Sigma _3$), the pair $(P,+)$ implies that we have a particle with positive
charge and the lower index $(+)$ denotes that the spin parity is positive
(upon action of operator $\beta $). With the usual eight canonical basis
vectors, ${\bf e}_i,i=1..8$ that are extensions of the two-dimensional KG
case, we can write the eight distinct possibilities for $\Psi $ as%
$$
u_{0\uparrow (+)}^{(P,+)}={\bf e}_1e^{-iE_p\tau /\hbar }\quad ;\quad
u_{0\downarrow (+)}^{(P,+)}={\bf e}_2e^{-iE_p\tau /\hbar };
$$
\begin{equation}
\label{59}u_{0\downarrow (+)}^{(A,+)}={\bf e}_3e^{+iE_p\tau /\hbar }\quad
;\quad u_{0\uparrow (+)}^{(A,+)}={\bf e}_4e^{+iE_p\tau /\hbar };
\end{equation}
$$
v_{0\uparrow (-)}^{(P,+)}={\bf e}_5e^{-iE_p\tau /\hbar }\quad ;\quad
v_{0\downarrow (-)}^{(P,+)}={\bf e}_6e^{-iE_p\tau /\hbar };
$$
\begin{equation}
\label{60}v_{0\downarrow (-)}^{(A,+)}={\bf e}_7e^{+iE_p\tau /\hbar }\quad
;\quad v_{0\uparrow (-)}^{(A,+)}={\bf e}_8e^{+iE_p\tau /\hbar },
\end{equation}
where
\begin{equation}
\label{61}E_p=m_0c^2.
\end{equation}

With the correspondence (55) between particle and antiparticle spinors, we
can write the spinors for $\Psi _{c1}$%
$$
\mu _{0\uparrow (+)}^{(A,-)}={\bf e}_1e^{+iE_p\tau /\hbar }\quad ;\quad \mu
_{0\downarrow (+)}^{(A,-)}={\bf e}_2e^{+iE_p\tau /\hbar };
$$
\begin{equation}
\label{62}\mu _{0\downarrow (+)}^{(P,-)}={\bf e}_3e^{-iE_p\tau /\hbar }\quad
;\quad \mu _{0\uparrow (+)}^{(P,-)}={\bf e}_4e^{-iE_p\tau /\hbar },
\end{equation}
$$
\nu _{0\uparrow (-)}^{(A,-)}={\bf e}_5e^{+iE_p\tau /\hbar }\quad ;\quad \nu
_{0\downarrow (-)}^{(A,-)}={\bf e}_6e^{+iE_p\tau /\hbar };
$$
\begin{equation}
\label{63}\nu _{0\downarrow (-)}^{(P,-)}={\bf e}_7e^{-iE_p\tau /\hbar }\quad
;\quad \nu _{0\uparrow (-)}^{(P,-)}={\bf e}_8e^{-iE_p\tau /\hbar },
\end{equation}
while for $\Psi _{c2}$
$$
\omega _{0\uparrow (-)}^{(A,-)}={\bf e}_5e^{+iE_p\tau /\hbar }\quad ;\quad
\omega _{0\downarrow (-)}^{(A,-)}={\bf e}_6e^{+iE_p\tau /\hbar };
$$
\begin{equation}
\label{64}\omega _{0\downarrow (-)}^{(P,-)}={\bf e}_7e^{-iE_p\tau /\hbar
}\quad ;\quad \omega _{0\uparrow (-)}^{(P,-)}={\bf e}_8e^{-iE_p\tau /\hbar
},
\end{equation}
$$
\eta _{0\uparrow (-)}^{(A,-)}={\bf e}_1e^{+iE_p\tau /\hbar }\quad ;\quad
\eta _{0\downarrow (-)}^{(A,-)}={\bf e}_2e^{+iE_p\tau /\hbar };
$$
\begin{equation}
\label{65}\eta _{0\downarrow (-)}^{(P,-)}={\bf e}_3e^{-iE_p\tau /\hbar
}\quad ;\quad \eta _{0\uparrow (-)}^{(P,-)}={\bf e}_4e^{-iE_p\tau /\hbar }.
\end{equation}

We also get the following relations between the antiparticle spinors
\begin{equation}
\label{66}\omega _{0\uparrow }=\nu _{0\downarrow \ };\omega _{0\downarrow
}=\nu _{0\uparrow }\mbox{ and }\eta _{0\uparrow }=\mu _{0\downarrow }\ ;\
\eta _{0\downarrow }=\mu _{0\uparrow };
\end{equation}
together with the spin parity relations
\begin{equation}
\label{67}\left\{
\begin{array}{c}
\beta u_0=+u_0 \\
\beta v_0=-v_0
\end{array}
\right. \ ;\ \left\{
\begin{array}{c}
\beta \mu _0=+\mu _0 \\
\beta \nu _0=-\nu _0
\end{array}
\right. \ ;\left\{
\begin{array}{c}
\beta \omega _0=-\omega _0 \\
\beta \eta _0=+\eta _0
\end{array}
\right. \ .
\end{equation}
These results can also be compared with those obtained using the linear
Dirac's equation\cite{7}.

We give the annihilation relations in Table 3.

We can now obtain the expression for the densities of probability and
current in the present formalism. This is a straightforward extension of
what was done in the KG formalism. We get equation
\begin{equation}
\label{68}\frac \partial {\partial t}\left( \Psi _{c2}\Psi \right) +\nabla
\cdot \left\{ \frac 1{2m}\left[ \Psi _{c2}\Lambda \nabla \Psi -\left( \nabla
\Psi _{c2}\right) \Lambda \Psi \right] -\frac{e{\bf A}}{mc}\Psi _{c2}\Lambda
\Psi \right\} =0,
\end{equation}
where
\begin{equation}
\label{69}\Lambda =\left( \Sigma _3+i\Sigma _2\right) .
\end{equation}
Equation (\ref{68}) then implies that
\begin{equation}
\label{70}\rho =\Psi _{c2}\Psi
\end{equation}
and
\begin{equation}
\label{71}{\bf j}=\frac 1{2m}\left[ \Psi _{c2}\Lambda \nabla \Psi -\left(
\nabla \Psi _{c2}\right) \Lambda \Psi \right] -\frac{e{\bf A}}{mc}\Psi
_{c2}\Lambda \Psi .
\end{equation}

We can write the expression for the density only in terms of $\Psi $
\begin{equation}
\label{72}\rho =\Psi ^{\dagger }\Sigma _3i\alpha _3\beta \Psi ,
\end{equation}
which gives, considering (\ref{68}), the conservation equation
\begin{equation}
\label{73}\frac{\partial \rho }{\partial t}+\nabla \cdot {\bf j}=0.
\end{equation}
With straightforward calculations we can show that one might write the
probability density, using the $\Psi $ components, as
\begin{equation}
\label{74}\rho =Im\left[ \sum_{ij}^2\varphi _i^{\dagger }\chi _j\right] .
\end{equation}

Finally, we shall comment the results of Table 3. Until now, we do not
impose any constraint on the parity of annihilating particles. This degree
of freedom leads to the possibility represented by the first column of Table
3, where particles and antiparticles with the same parity annihilate each
other. The problem with this annihilation process is that physicists have
not found, until now, any spin zero (longitudinal) photon. There are indeed
strong arguments against their existence. We could then postulate that
particles and antiparticles should have the spin parity (if any) also
reverted. However, we avoid to assert this for the moment, since we are here
interested in uncovering worlds, not in hiding them.

\section{Conclusions}

We have then succeeded in showing that our conjecture can be accommodated
into the formal apparatus of the special relativistic quantum mechanics.

When Dirac's theory, based on its first order equation, revealed the
antiparticle (as were defined), many physicists were delighted with the
symmetries it has brought\cite{4,8}. For each element with mass of a given
value, Nature does not distinguish them by giving different charges. So, the
massive proton with positive charge shall have its negative charge
counterpart. Each particle has its antiparticle defined by its charge mirror.

This work takes this approach to its utmost limits. Each particle has its
antiparticle defined by its mirror world, where both charge and mass signs
are reverted. Also, there is no particle being its own antiparticle in the
sense that only entities with opposite mass sign might annihilate each
other. In this case, pions zero are annihilated by antipions zero (both, of
course, might decay spontaneously).

The vacuum that emerges is not a filled structure in which every point of
the real space is occupied with an infinitude of antiparticles. This picture
can be avoided while keeping the important property of vacuum polarization.
Moreover, contrarily to the usual interpretation, the present theory treats
particles and antiparticles in a totally symmetrical way $^{(5)}$. We shall
also stress, considering the present conjecture, that the gravitational
field is highly capable of polarizing the vacuum. This property will become
relevant in the second paper of this series.

This theory does not claim for a strict inertial mass conservation law. This
is because for mass we have Einstein's equation, $E=mc^2$, which
distinguishes mass from charge with respect to conservative behavior. If we
also admit, following the discussion at the end of the last section, that
creation and annihilation processes shall conserve parity, then we place
parity, aside from the charge, as a fundamental property of Nature.

The possible existence of negative masses have far reaching cosmological
consequences that will be addressed in a future paper.

The arguments above, about the higher symmetry of Nature introduced by the
concept of negative masses, cannot, of course, prove the conjecture. They
fail to have any necessity character. They are just a metaphysical
constraint we wish to impose upon Nature. The final word will be with the
experimental physicists. This formidable task is being presently carried on
by several experiments\cite{9}.

Clearly, in the realm of special relativistic quantum mechanics, fixing mass
signs is an ad hoc postulate, as we stated in the last paragraph. The next
paper of this series will show, however, that the general relativistic
quantum mechanical theory derived in paper II of this series supports this
conjecture.

\appendix

\section{Negative Densities}

When studying the KG formalism, we are faced with a striking fact. While the
amplitudes in (\ref{7}) indicate that we should expect both positive and
negative energy densities, the energy density obtained from the
energy-momentum tensor is always positive. Moreover, since we sustain\cite{%
1,2,3} that relativistic quantum mechanics can be derived from
classical relativity and statistics, where we impose the positive character
of the energy, it is higly desirable to clarify this apparent paradox.

We will show in this appendix that this paradoxical situation can be easily
clarified. We will use the formalism already developed\cite{1,2,3} that
enables us to go from Liouville's equation to the equation for the density
function. From the analysis of what is happening in phase space, it will be
easier to understand this property of the KG equation. In fact, it will be
shown that this ''pathology'' is also present in the non-relativistic
Schr\"odinger equation. We will thus present both the non-relativistic and
relativistic calculations to make our discussion clearer.

We have shown\cite{1,2,3} that all non-relativistic and relativistic
quantum mechanics could be obtained from the classical Liouville's equation
\begin{equation}
\label{75}\frac{dF_n\left( {\bf x},{\bf p};t\right) }{dt}=0\ ;\ \frac{%
dF_r\left( x,p\right) }{d\tau }=0,
\end{equation}
where ${\bf x}$ and ${\bf p}$ are the position and momentum vectors, $x$ and
$p$ are the related four-vectors, $\tau $ is the proper time and $F_n$ and $%
F_r$ are the non-relativistic and relativistic joint probability densities,
respectively. This was accomplished using the Infinitesimal Wigner-Moyal
Transformations
\begin{equation}
\label{76}\rho _n^{(d)}\left( {\bf x-}\frac{\delta {\bf x}}2,{\bf x}+\frac{%
\delta {\bf x}}2;t\right) =\int F_n\left( {\bf x},{\bf p};t\right) \exp
\left( \frac i\hbar {\bf p}\cdot \delta {\bf x}\right) d^3p
\end{equation}
and
\begin{equation}
\label{77}\rho _r^{(d)}\left( x{\bf -}\frac{\delta x}2,x+\frac{\delta x}%
2\right) =\int F_r\left( x,p;t\right) \exp \left( \frac i\hbar p^\alpha
\delta x_\alpha \right) d^4p,
\end{equation}
where $\rho _n$ and $\rho _r$ are the non-relativistic and relativistic
density functions, respectively. We also assumed as an axiom that Newton's
equation, and its special relativistic counterpart, are valid
\begin{equation}
\label{78}\frac{d{\bf x}}{dt}=\frac{{\bf p}}m\ ;\ \frac{dx^\alpha }{d\tau }=
\frac{p^\alpha }m;
\end{equation}
to obtain the equations
\begin{equation}
\label{79}\frac{-\hbar ^2}{2m}\frac{\partial ^2\rho _n^{(d)}}{\partial {\bf x%
}\partial \left( \delta {\bf x}\right) }=i\hbar \frac{\partial \rho _n^{(d)}
}{\partial t}\ ;\ \hbar ^2\frac{\partial ^2\rho _r^{(d)}}{\partial x^\alpha
\partial \left( \delta x_\alpha \right) }=0.
\end{equation}
These equations, we showed, can be taken into the Schr\"odinger's and
Klein-Gordon's equations (in the absence of external forces and spin, for
simplicity)
\begin{equation}
\label{80}\frac{-\hbar ^2}{2m}\frac{\partial ^2\psi _n}{\partial {\bf x}^2}%
=i\hbar \frac{\partial \psi _n}{\partial t}\ ;\ \left( \hbar ^2\Box
-m^2\right) \psi _r=0,
\end{equation}
where $\psi _n$ and $\psi _r$ are the non-relativistic and relativistic
probability amplitudes, when we use the property that $\delta x$ represents
an infinitesimal variation and that the expansions
\begin{equation}
\label{81}\rho _{n(+)}^{(d)}\left( {\bf x-}\frac{\delta {\bf x}}2,{\bf x}+
\frac{\delta {\bf x}}2;t\right) =\psi _n^{*}\left( {\bf x-}\frac{\delta {\bf %
x}}2;t\right) \psi _n\left( {\bf x}+\frac{\delta {\bf x}}2;t\right)
\end{equation}
and
\begin{equation}
\label{82}\rho _{r(+)}^{(d)}\left( x{\bf -}\frac{\delta x}2,x+\frac{\delta x}%
2\right) =\psi _r^{*}\left( x{\bf -}\frac{\delta x}2\right) \psi _r\left( x+
\frac{\delta x}2\right)
\end{equation}
might be performed.

{}From these expressions, we can define the 3 and 4-momentum operators by
means of the expressions for their expectation values
\begin{equation}
\label{83}\left\langle {\bf p}\right\rangle =\lim _{\delta {\bf x}%
\rightarrow 0}\frac \hbar i\frac \partial {\partial \left( \delta {\bf x}%
\right) }\int \rho _{n(+)}^{(d)}\left( {\bf x-}\frac{\delta {\bf x}}2,{\bf x}%
+\frac{\delta {\bf x}}2;t\right) d^3x;
\end{equation}
\begin{equation}
\label{84}\left\langle p\right\rangle =\lim _{\delta x\rightarrow 0}\frac
\hbar i\frac \partial {\partial \left( \delta x\right) }\int \rho
_{r(+)}^{(d)}\left( x{\bf -}\frac{\delta x}2,x+\frac{\delta x}2;t\right)
d^4x.
\end{equation}

It is noteworthy that we have, however, a freedom of choice in expressions (%
\ref{81}) and (\ref{82}). We could equally well have chosen
\begin{equation}
\label{85}\rho _{n(-)}^{(d)}\left( {\bf x-}\frac{\delta {\bf x}}2,{\bf x}+
\frac{\delta {\bf x}}2;t\right) =\psi _n^{*}\left( {\bf x+}\frac{\delta {\bf %
x}}2;t\right) \psi _n\left( {\bf x-}\frac{\delta {\bf x}}2;t\right) =\rho
_{n(+)}^{(d)\dagger }
\end{equation}
and
\begin{equation}
\label{86}\rho _{r(-)}^{(d)}\left( x{\bf -}\frac{\delta x}2,x+\frac{\delta x}%
2\right) =\psi _r^{*}\left( x+\frac{\delta x}2\right) \psi _r\left( x-\frac{%
\delta x}2\right) =\rho _{r(+)}^{(d)\dagger }
\end{equation}
that is equivalent to the change $\psi \leftrightarrow \psi ^{*},i=n,r$. It
is easy to see that, with this new definition, we get
\begin{equation}
\label{87}\left\langle {\bf p}\right\rangle \rightarrow -\left\langle {\bf p}%
\right\rangle \ ;\ \left\langle p\right\rangle \rightarrow -\left\langle
p\right\rangle .
\end{equation}
We can interpret these results as representing, in the non-relativistic
case, a problem where the particle travels back in space. In the
relativistic case it can be understood as if the particle travels back in
space-time (with negative momentum and energy).

However, if we still want to have an adequate definition of three and
four-momentum, as given by (\ref{83}) and (\ref{84}), we shall redefine the
3- and 4-momentum mean values as
\begin{equation}
\label{88}\left\langle {\bf p}\right\rangle =\lim _{\delta {\bf x}%
\rightarrow 0}-\frac \hbar i\frac \partial {\partial \left( \delta {\bf x}%
\right) }\int \rho _{n(-)}^{(d)}\left( {\bf x-}\frac{\delta {\bf x}}2,{\bf x}%
+\frac{\delta {\bf x}}2;t\right) d^3x;
\end{equation}
\begin{equation}
\label{89}\left\langle p\right\rangle =\lim _{\delta x\rightarrow 0}-\frac
\hbar i\frac \partial {\partial \left( \delta x\right) }\int \rho
_{r(-)}^{(d)}\left( x{\bf -}\frac{\delta x}2,x+\frac{\delta x}2;t\right)
d^4x.
\end{equation}
which give the operators
\begin{equation}
\label{90}{\bf p}_{op}=\lim _{\delta {\bf x}\rightarrow 0}-\frac \hbar
i\frac \partial {\partial \left( \delta {\bf x}\right) }\ ;\ p_{op}=\lim
_{\delta x\rightarrow 0}-\frac \hbar i\frac \partial {\partial \left( \delta
x\right) }.
\end{equation}
In general, we have
\begin{equation}
\label{91}{\bf p}_{op}=\lim _{\delta {\bf x}\rightarrow 0}\lambda \frac
\hbar i\frac \partial {\partial \left( \delta {\bf x}\right) }\ ;\
p_{op}=\lim _{\delta x\rightarrow 0}\lambda \frac \hbar i\frac \partial
{\partial \left( \delta x\right) },
\end{equation}
where $\lambda $ has the same definition given in the main text, when acting
upon $\rho _{i,\lambda },i=n,r$. This assures that the energy and momentum
have the correct sign when calculated by expressions (\ref{83}-\ref{84}) or (%
\ref{88}-\ref{89}).

It is important to stress that equations (\ref{79}) do not depend upon $%
\lambda $, since they are quadratic in the considered quantities (energy and
momentum for KG's and momentum for Schr\"odinger's). This justifies that the
energy density obtained from the energy-momentum tensor is always positive.
In the same way, the energy in Schr\"odinger's equation is not affected,
since the time does not enter into the non-relativistic transformation (\ref
{76}).

With the conventions (\ref{91}), the relativistic energy density\cite{1,2,3}
can be written, in the absence of electromagnetic fields, as
\begin{equation}
\label{92}p_\lambda ^0\left( x\right) =\frac{i\hbar }2\lambda \left[ \psi
^{*}\frac{\partial \psi }{\partial t}-\psi \frac{\partial \psi ^{*}}{%
\partial t}\right] =\frac{i\hbar }2\left[ \psi ^{*}\frac{\partial \psi }{%
\partial t}-\psi \frac{\partial \psi ^{*}}{\partial t}\right] _{+},
\end{equation}
which is always positive. If we impose the possibility of negative masses
for the complex conjugate amplitudes, the probability density can be written
as
\begin{equation}
\label{93}\rho _\lambda \left( x\right) =\frac{i\hbar }{2\lambda mc^2}\left[
\psi ^{*}\frac{\partial \psi }{\partial t}-\psi \frac{\partial \psi ^{*}}{%
\partial t}\right] _{+},
\end{equation}
where $\lambda $ now comes from the mass sign.

Let us consider now the relativistic situation when electromagnetic fields
are present. The energy density is now given by
\begin{equation}
\label{94}p_\lambda ^0\left( x\right) =\frac{i\hbar }2\left[ \psi ^{*}\frac{%
\partial \psi }{\partial t}-\psi \frac{\partial \psi ^{*}}{\partial t}%
\right] _{+}-\lambda e\Phi \psi ^{*}\psi ,
\end{equation}
where $e$ is the particle charge and $\Phi $ is the scalar electromagnetic
potential. The parameter $\lambda $ appears multiplying the charge, since in
the equation that $\varphi ^{*}$ solves, the charge changes sign. The
probability density can be written as
\begin{equation}
\label{95}\rho _\lambda \left( x\right) =\frac{i\hbar }{2\lambda mc^2}\left[
\psi ^{*}\frac{\partial \psi }{\partial t}-\psi \frac{\partial \psi ^{*}}{%
\partial t}\right] _{+}-\frac e{mc^2}\Phi \psi ^{*}\psi ,
\end{equation}
as was presented in the main text.

It is important to note that (\ref{95}) is a density that represents
particles (positive mass, given by $\psi $) and antiparticles (negative
mass, given by $\psi ^{*}$) with energy and momentum positive, as stressed
in the main text (\ref{9}). We thus obtain results in agreement with those
obtained with the energy-momentum tensor.

The final expression for the density with the normalization
\begin{equation}
\label{96}\int \rho _\lambda \left( x\right) d^3x=\lambda
\end{equation}
implies that, without fields, the worlds of particles and antiparticles fall
apart. This is a very important property; it allows us to avoid the vacuum
picture that emerges from Dirac's theory based in his first order equation.
This theory is automatically prevented from a radiative catastrophe.

The above calculations also allow us to understand the picture of particle
flow. A positive mass particle, with negative energy and momentum traveling
backward on time, is equivalent to a negative mass antiparticle, with
positive energy and momentum, traveling in the usual time direction. These
conventions just reflect equations (\ref{10}-\ref{12}) and agree with the
signs of the velocity as being opposite for particles and antiparticles.
These considerations will play an important role in the next paper of this
series.

\newpage

\begin{table}
 \begin{center}
   \begin{tabular}{|c|c|c|} \hline
       {\bf mass}  &  {\bf charge}  &  {\bf amplitude}   \\  \hline
             +         &          +          &
             $\chi_1$          \\ \hline
             +         &          -          &
             $\chi_{2}^{\dag}$ \\ \hline
             -          &          -          &
             $\chi_{1}^{\dag}$ \\ \hline
             -          &          +         &
             $\chi_2$          \\ \hline
  \end{tabular}
 \end{center}
 \caption{Possible combinations of mass and charge signals allowed by Nature
 for spinless particles.}
\end{table}

\newpage

\begin{table}
 \begin{center}
   \begin{tabular}{|c|c|c|c|} \hline
       {\bf mass}  &  {\bf charge}  &  {\bf spin}         & {\bf amplitude}
     \\  \hline
             +         &          +          &
             $\uparrow$     &   $\phi_1$                \\ \hline
             +         &          +          &
             $\downarrow$ &   $\chi_{1}$              \\ \hline
             +         &          -          &
             $\uparrow$     &   $\chi_{2}^{\dag}$    \\ \hline
             +         &          -          &
             $\downarrow$ &   $\phi_{2}^{\dag}$    \\ \hline
             -          &          +          &
             $\uparrow$     &   $\phi_{2}$               \\ \hline
             -          &          +         &
             $\downarrow$ &   $\chi_2$                  \\ \hline
             -          &          -          &
             $\uparrow$     &   $\chi_{1}^{\dag}$     \\ \hline
             -          &          -          &
             $\downarrow$ &   $\phi_{1}^{\dag}$     \\ \hline
  \end{tabular}
 \end{center}
 \caption{Possible combinations allowed by Nature for particles with spin.}
\end{table}

\newpage

\begin{table}
 \begin{center}
   \begin{tabular}{|c|c|c|} \hline
       ${\bf \Psi}$                  &  ${\bf \Psi}_{c1}(\hbar\omega)$  &
       ${\bf \Psi}_{c2}(\hbar\omega)$             \\  \hline
$u_{0\uparrow(+)}^(P,+)$      &  $\mu_{0\downarrow(+)}^{(A,-)}(0)$   &
$\omega_{0\uparrow(-)}^{(A,-)}(+1)$      \\ \hline
$u_{0\downarrow(+)}^(P,+)$  &  $\mu_{0\uparrow(+)}^{(A,-)}(0)$       &
$\omega_{0\downarrow(-)}^{(A,-)}(-1)$  \\ \hline
$u_{0\downarrow(+)}^(A,+)$  &  $\mu_{0\uparrow(+)}^{(P,-)}(0)$       &
$\omega_{0\downarrow(-)}^{(P,-)}(-1)$  \\ \hline
$u_{0\uparrow(+)}^(A,+)$      &  $\mu_{0\downarrow(+)}^{(P,-)}(0)$   &
$\omega_{0\uparrow(-)}^{(P,-)}(+1)$      \\ \hline\hline
$v_{0\uparrow(-)}^(P,+)$       &  $\nu_{0\downarrow(-)}^{(A,-)}(0)$   &
$\eta_{0\uparrow(+)}^{(A,-)}(+1)$           \\ \hline
$v_{0\downarrow(-)}^(P,+)$   &  $\nu_{0\uparrow(-)}^{(A,-)}(0)$       &
$\eta_{0\downarrow(+)}^{(A,-)}(-1)$       \\ \hline
$v_{0\downarrow(-)}^(A,+)$   &  $\nu_{0\uparrow(-)}^{(P,-)}(0)$       &
$\eta_{0\downarrow(+)}^{(P,-)}(-1)$      \\ \hline
$v_{0\uparrow(-)}^(A,+)$       &  $\nu_{0\downarrow(-)}^{(P,-)}(0)$   &
$\eta_{0\uparrow(+)}^{(P,-)}(+1)$      \\ \hline
  \end{tabular}
 \end{center}
 \caption{Possible combinations allowed by Nature for particles with spin.}
\end{table}

\newpage

\unitlength=1.00mm
\special{em:linewidth 1pt}
\linethickness{1pt}

\begin{figure}
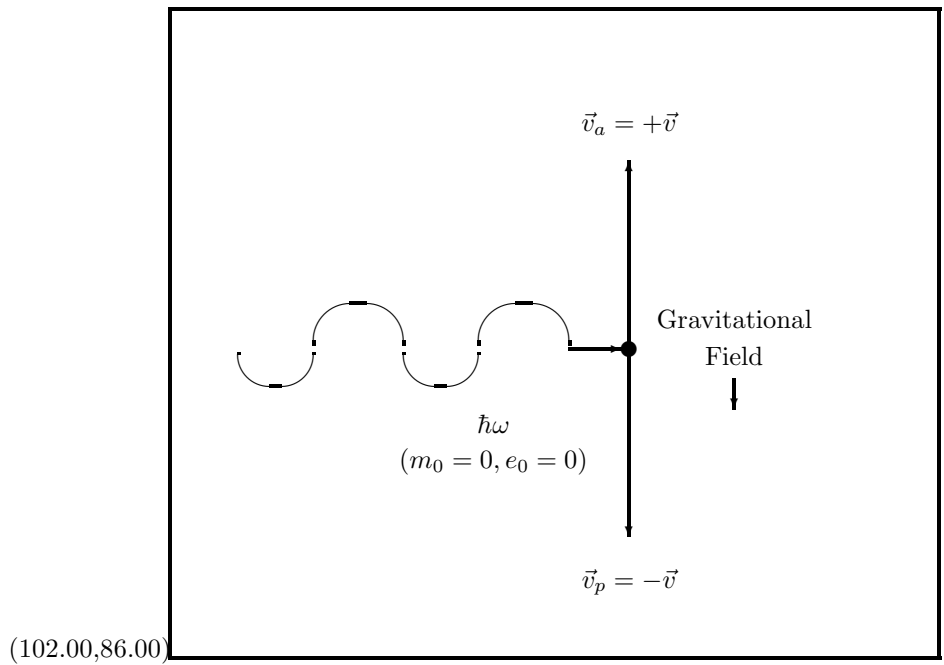
(102.00,86.00)
\put(61.00,41.00){\circle*{2.00}}
\put(61.00,42.00){\vector(0,1){24.00}}
\put(61.00,40.00){\vector(0,-1){24.00}}
\put(53.00,41.00){\vector(1,0){7.00}}
\put(43.00,31.00){\makebox(0,0)[cc]{$\hbar\omega$}}
\put(43.00,26.00){\makebox(0,0)[cc]{$(m_0=0,e_0=0)$}}
\put(61.00,10.00){\makebox(0,0)[cc]{$\vec{v}_p=-\vec{v}$}}
\put(61.00,71.00){\makebox(0,0)[cc]{$\vec{v}_a=+\vec{v}$}}
\put(75.00,45.00){\makebox(0,0)[cc]{Gravitational}}
\put(75.00,40.00){\makebox(0,0)[cc]{Field}}
\put(75.00,37.00){\vector(0,-1){4.00}}
\put(14.00,40.50){\oval(10.00,9.00)[b]}
\put(25.00,41.50){\oval(12.00,11.00)[t]}
\put(36.00,40.50){\oval(10.00,9.00)[b]}
\put(47.00,41.50){\oval(12.00,11.00)[t]}
\put(0.00,0.00){\framebox(102.00,86.00)[cc]{}}
\caption{Particle-antiparticle trajectories in the presence of a
gravitational field.}
\end{figure}

\newpage

\begin{figure}
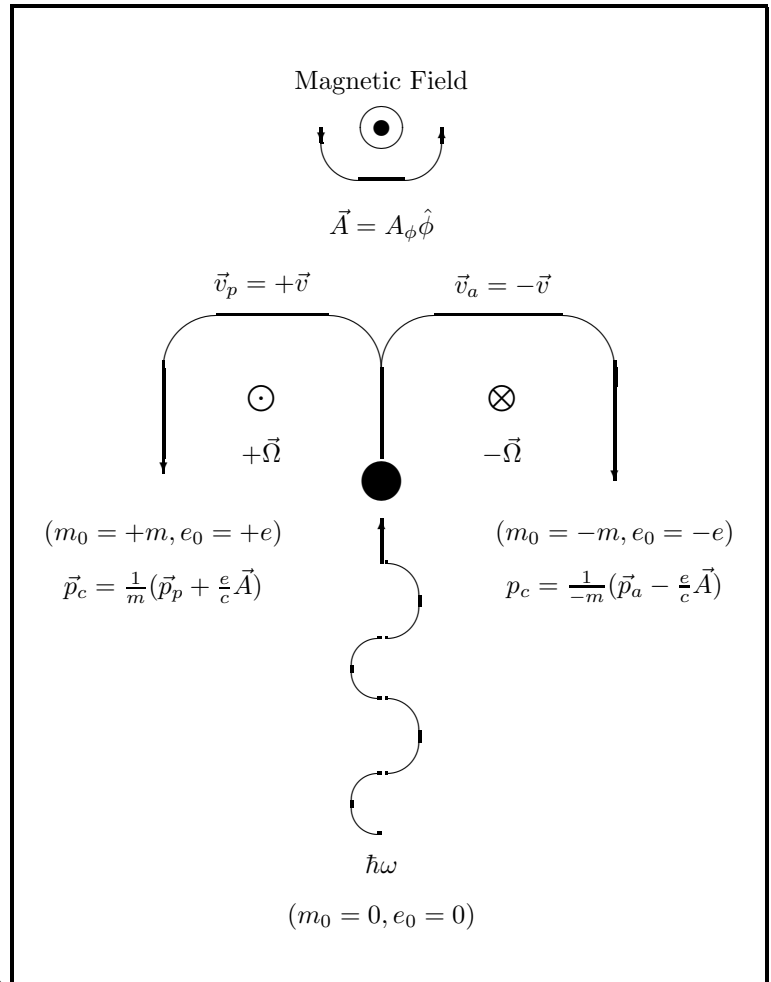
(101.00,132.00)
\put(50.00,26.00){\oval(8.00,8.00)[l]}
\put(50.50,35.00){\oval(9.00,10.00)[r]}
\put(50.00,44.00){\oval(8.00,8.00)[l]}
\put(50.50,53.00){\oval(9.00,10.00)[r]}
\put(50.00,58.00){\vector(0,1){6.00}}
\put(50.00,69.00){\circle*{5.20}}
\put(35.50,72.00){\oval(29.00,38.00)[t]}
\put(65.50,81.50){\oval(31.00,19.00)[t]}
\put(21.00,85.00){\vector(0,-1){15.00}}
\put(81.00,85.00){\vector(0,-1){16.00}}
\put(66.00,80.00){\makebox(0,0)[cc]{$\bigotimes$}}
\put(81.00,62.00){\makebox(0,0)[cc]{$(m_0=-m,e_0=-e)$}}
\put(81.00,55.00){\makebox(0,0)[cc]{$p_c=\frac{1}{-m}(\vec{p}_a
-\frac{e}{c}\vec{A})$}}
\put(34.00,80.00){\makebox(0,0)[cc]{$\bigodot$}}
\put(21.00,62.00){\makebox(0,0)[cc]{$(m_0=+m,e_0=+e)$}}
\put(21.00,55.00){\makebox(0,0)[cc]{$\vec{p}_c=\frac{1}{m}(\vec{p}_p
+\frac{e}{c}\vec{A})$}}
\put(50.00,18.00){\makebox(0,0)[cc]{$\hbar\omega$}}
\put(50.00,11.00){\makebox(0,0)[cc]{$(m_0=0,e_0=0)$}}
\put(34.00,95.00){\makebox(0,0)[cc]{$\vec{v}_p=+\vec{v}$}}
\put(66.00,95.00){\makebox(0,0)[cc]{$\vec{v}_a=-\vec{v}$}}
\put(50.00,122.00){\makebox(0,0)[cc]{Magnetic Field}}
\put(50.00,116.00){\circle{6.00}}
\put(50.00,116.00){\circle*{2.00}}
\put(50.00,114.00){\oval(16.00,10.00)[b]}
\put(58.00,114.00){\vector(0,1){2.00}}
\put(42.00,116.00){\vector(0,-1){2.00}}
\put(50.00,103.00){\makebox(0,0)[cc]{$\vec{A}=A_{\phi}\hat{\phi}$}}
\put(34.00,73.00){\makebox(0,0)[cc]{$+\vec{\Omega}$}}
\put(66.00,73.00){\makebox(0,0)[cc]{$-\vec{\Omega}$}}
\put(1.00,2.00){\framebox(100.00,130.00)[cc]{}}
\caption{Particle-Antiparticle trajectories in the presence of a
homogeneous magnetic field.}
\end{figure}

\end{document}